# Image Compression and Decompression Framework Based on Latent Diffusion Model for Breast Mammography

InChan Hwang
School of Data Science & Analytics
Kennesaw State University
Kennesaw, GA, USA
ihwang@students.kennesaw.edu

MinJae Woo
School of Data Science & Analytics
Kennesaw State University
Kennesaw, GA, USA
mwoo1@kennesaw.edu

*Abstract*—This research presents a novel framework for the compression and decompression of medical images utilizing the Latent Diffusion Model (LDM). The LDM represents advancement over the denoising diffusion probabilistic model (DDPM) with a potential to yield superior image quality while requiring fewer computational resources in the image decompression process. A possible application of LDM and Torchvision for image upscaling has been explored using medical image data, serving as an alternative to traditional image compression and decompression algorithms. The experimental outcomes demonstrate that this approach surpasses a conventional file compression algorithm, and convolutional neural network (CNN) models trained with decompressed files perform comparably to those trained with original image files. This approach also significantly reduces dataset size so that it can be distributed with a smaller size, and medical images take up much less space in medical devices. The research implications extend to noise reduction in lossy compression algorithms and substitute for complex wavelet-based lossless algorithms.

## I. INTRODUCTION

People have historically engaged in the creation of images across various domains, encompassing painting, photography, and medical examinations. Decision-making processes oftentimes heavily rely on the evidence and information conveyed through visual materials. Similarly, professionals like radiologists analyze medical images to identify features such as carcinoma in-situ, architectural distortion, breast density, and calcification, crucial for determining optimal cancer treatment strategies [1]. To facilitate the clinical decision making process, both medical imaging and computer vision researchers have advocated for computer-aided diagnosis systems powered by AI/ML models to enhance diagnostic accuracy [1, 2]. Some researchers have devised models leveraging image data exclusively or in conjunction with clinical data for disease diagnosis [3, 4].

Computer imaging modalities yield high-resolution images with a high bit-depth, resulting in image data that holds abundant information, particularly when human organs and tissues are scanned using radiographic imaging techniques [5]. The Digital Imaging and Communications in Medicine (DICOM) format has seen widespread adoption across various medical devices, effectively catering to this specific imaging need [6].

The field of medical imaging holds significant importance due to its critical role in enabling accurate diagnosis of patients' conditions, necessitating careful processing of imaging data. Operating medical imaging devices raises two prominent challenges. Medical images are captured at high resolutions, resulting in exceedingly large file sizes. Consequently, medical devices often face constraints in file storage space. Transmitting such massive image data files over the Internet is time-consuming. File compression techniques are commonly employed to mitigate transfer times for all files, including medical images. However, compression may lead to information loss in media files. Researchers in the field of medical imaging and related disciplines actively pursue compression algorithms that ensure data integrity while minimizing their sizes after the compression [7]. Another challenge involves the removal of noise in acquired medical images, which has been a focal area of research in the field of medical imaging [8].

Image data compression and decompression are crucial processes aimed at reducing disk storage consumption and minimizing network bandwidth usage [9]. Compression involves the removal or encoding of data to decrease file sizes, thus enhancing storage efficiency and facilitating faster data transmission over networks. Two fundamental approaches to image data compression exist: lossy and lossless. The lossy approach entails the removal of pixel information to achieve compression, albeit at the cost of losing some data. On the other hand, the lossless approach involves encoding pixel data while preserving all information without any loss, ensuring accurate reconstruction upon decompression [7, 10].

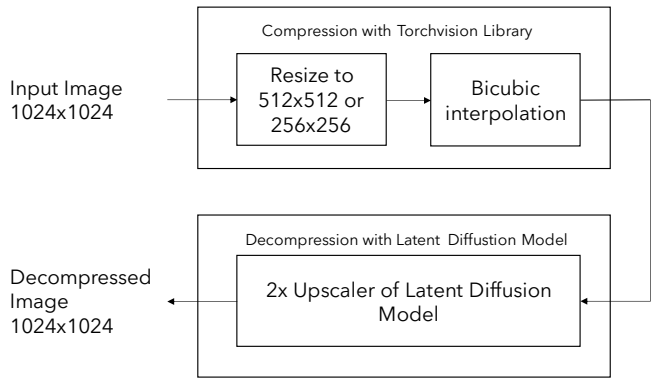

**Figure 1. LDM and Torchvision Medical Image Compression Framework**

Following the advent of Convolutional Neural Network (CNN) in image data analysis, deep learning (DL) models efficiently extract distinctive image features, allowing for unique image identification with a reduced size. This characteristic of CNN presents a superior alternative to traditional lossy image compression algorithms like JPEG. CNN excels at solving non-linear problems by training numerous neural layers and optimizing both the encoder and decoder modules within the neural layers. Consequently, CNN-based compression models have significantly bolstered model efficiency in image compression applications [11].

In this study, we propose a novel framework for breast image compression and decompression utilizing the LDM and Torchvision Vision Transformer as the decoder and encoder components respectively. Notably, this research marks the proactive attempt utilizing the framework of LDM and Torchvision library to compress and decompress breast medical images, offering an in-depth analysis of performance degradation in decompressed images. We also explore the effectiveness of the CNN model when trained and validated using these decompressed images, presenting a comprehensive evaluation of this novel approach.

## II. RELATED WORKS

### A. Generative Adversarial Networks (GAN)

Generative Adversarial Networks (GANs) are highly effective in producing high-quality synthetic images. A GAN consists of two integral neural network components: the generator and the discriminator modules. The generator module is responsible for generating fake images by randomly sampling from the image distribution learned from the training set. Conversely, the discriminator module evaluates whether an image is real from the image distribution or fake generated by the generator module. These modules undergo simultaneous training to optimize their respective functionalities. Ultimately, the generator module is trained to calculate the parameters of the image distribution of real image data [14]. However, it is important to note that optimizing and training the GAN model is considerably more demanding compared to other generative models [12].

### B. Diffusion Models

The Denoising Diffusion Probabilistic Model (DDPM) has demonstrated superior image distribution estimation and sample image quality compared to Generative Adversarial Networks (GANs) when drawing images from the source image distribution [15]. Within DDPM, the integration of U-Net helps estimate noise parameter distributions within the forwarded noised image data, effectively denoising noised images in the reverse process. Synthesized images achieve optimal quality when utilizing a modified objective function during the training phase [16]. However, DDPM variations exhibit drawbacks in terms of low inference performance and high computing resource consumption. Addressing these concerns, LDM represents an enhancement over DDPM, effectively compressing high-dimensional data into a lower image dimension. This advancement significantly reduces inference time and computing resource consumption while preserving the high image quality of synthesized images [12].

### C. ResNet Models

In the realm of Neural Networks, models with a greater number of layers exhibit enhanced performance compared to those with a small number of layers, while operating with a limited set of parameters. However, as the number of convolution layers increases, the process of updating gradients during backpropagation becomes less effective due to some pixel values converging to zero. This issue, known as the vanishing gradient problem, hampers the effective flow of gradients across the network. To mitigate this challenge, the Residual Neural Network (ResNet) was introduced, incorporating skip connections between residual blocks. These skip connections enable elementwise addition between the input and output of a residual block, ensuring that gradients are preserved and preventing the vanishing gradient problem. Consequently, off-the-shelf ResNet architectures often can accommodate a maximum of 152 layers, highlighting its effectiveness in addressing this key limitation [17].

### D. Torchvision library

Torchvision is an open-source software for computer vision, distributed under the BSD license, and it has gained widespread adoption within the Machine Learning (ML) and Artificial Intelligence (AI) communities. Serving as a superset library, PyTorch offers AI and ML algorithms with a user-friendly interface, encompassing a range of models such as Hidden Markov Models, Support Vector Machines, and Gaussian Mixture Models. Within this ecosystem, Torchvision operates as a subset library, specifically focusing on software interfaces related to image manipulation and image processing techniques. Notably, both PyTorch and Torchvision share identical image datatypes, allowing for seamless integration and the direct feeding of processed image data from Torchvision into ML/AI models constructed using PyTorch [13].

**Table 1. EMBED Dataset Information**

| Dataset | Size | Clinical Information | Release Year | Resolution |
|---|---|---|---|---|
| EMBED | 364,791 images | Age, tissue density, mass shapes, image findings, BIRAD No., pathological outcomes | 2023 | 2294x1914 ~ 4096 x3328 |

*E. Emory Breast Imaging Dataset (EMBED)*

The publicly released EMBED dataset comprises complete field digital mammograms of 23,264 patients, encompassing a total of 364,791 images. Notably, the dataset exhibits a distinctive racial demographic composition compared to other publicly available breast mammography datasets, incorporating 39% White, 42% Black, and 6% Asian populations. Furthermore, this dataset provides comprehensive clinical information, including patient ages, tissue density, mass shapes, image findings, imaging descriptors, Breast Imaging-Reporting and Data System (BIRAD) scores, and pathological outcomes. Since its release in 2023, this dataset has been made accessible to the public [18]. Its brief information is depicted in Table 1.

## III. METHODOLOGY

The preprocessing of the EMBED dataset requires domain specific knowledge. This section outlines the labeling process, categorizing mammograms with abnormality as "Positive" and those without as "Negative" as guided by recent relevant works on this dataset [19]. The preprocessing is followed by compression and subsequent decompression, essential for minimizing file sizes and restoring files to their original states. Next, Customized ResNet models are trained and validated using the source images. The models are also trained and validated using the decompressed images, allowing for a comparative performance analysis to understand the distinctions between these two model types.

*A. Data preprocessing and methods*

The EMBED dataset initially consists of images in DICOM format. To enhance accessibility and usability, these images are converted into 16-bit grayscale PNG format. Additionally, their resolutions are adjusted to 1024x1024 using bicubic interpolation and anti-aliasing, ensuring they fit into GPU memory while preserving crucial information [20]. The selection of positive cases involves images categorized with BIRAD scores 4, 5, and 6 of the patients for these images who undergo a biopsy within 180 days. Conversely, negative cases are sourced from images categorized with BIRAD scores 1 and 2. The assignment of BIRAD numbers to each mammogram is based on clinical findings reported by radiologists. Subsequently, train, validation, and test datasets are partitioned using a 60:20:20 ratio, preventing any patient leakage between these datasets [19], as illustrated in Table 2.

These preprocessed images serve as the source images for further analysis and model training.

*B. Compression and Decompression*

The preprocessed mammograms, sized at 1024x1024, undergo a compression procedure using the Torchvision transform function. This compression operation effectively transforms the images into two distinct sets: one set at a resolution of 512x512 and another at 256x256. The resizing process incorporates bicubic interpolation for both resolutions, ensuring precise and high-quality transformations crucial for subsequent analysis and classification in the mammography domain. To optimize compression performance across a multitude of images, multithreaded parallelization is implemented around the image transform function. Specifically, a total of four threads are allocated to efficiently rescale the entire dataset, including training, validation, and test set images. This entire process represents the compression module.

The decompression of resized images is achieved using the LDM upscaler module, which is loaded with the "sd-x2-latent-upscaler" pretrained model. This model is specifically designed to upscale images to twice their original resolution, effectively transforming 512x512 resolution images to 1024x1024. In the upscaling process, a total of 20 inference steps are performed to accurately upscale the images, with a guidance scale for inference set to 0. Additionally, upscaling 256x256 resolution images to 1024x1024 involves a recursive process utilizing "sd-x2-latent-upscaler," applied twice to the 256x256 resolution images, resulting in the creation of original-resolution 1024x1024 images. The compression and subsequent decompression processes are visually illustrated in Figure 1.

*C. ResNet50 Breast Cancer Classifier*

In the design of the breast cancer classifier, a customized ResNet50 architecture has been selected. The integration of skip connections between residual blocks serves to enrich the classifier model with an abundance of neural layers. This design choice allows for the extraction of diverse image features throughout the convolution layers, ultimately enhancing the model's ability to discern intricate patterns within the image data. The ResNet50 architecture has garnered significant attention in the field of cancer classification as well as breast cancer, and has been widely adopted in numerous research endeavors aiming to develop robust and accurate classification models [19].

**Table 2. Dataset Configuration for Train, Validation, and Test**

| Image Label | Total Image Count | Train | Validation | Test |
|---|---|---|---|---|
| Positive | 1,187 | 713 | 237 | 237 |
| Negative | 1,187 | 713 | 237 | 237 |

**Table 3. Compression Performance Comparison**

| | Source 1024x1024 | Compressed to 512x512 | Compressed to 256x256 |
|---|---|---|---|
| **LDM/Torchvision Compression** | 564 MB | 266MB | 81MB |
| **ZIP compression** | 577 MB | | |

Table 4. ResNet Model Performance on Test Datasets

| Image Resolution | AUC | Accuracy | Precision | Recall |
|---|---|---|---|---|
| Source and decompressed 1024x1024 images from 512x512 and 256x256 images | | | | |
| 1024x1024 | 0.79 (0.75 – 0.82) | 0.75 (0.73 – 0.78) | 0.73 (0.69 – 0.77) | 0.78 (0.75 – 0.82) |
| 1024x1024 [512x512] | 0.78 (0.74 – 0.81) | 0.74 (0.72 – 0.77) | 0.75 (0.70 – 0.79) | 0.72 (0.67 – 0.77) |
| 1024x1024 [256x256] | 0.74 (0.71 – 0.77) | 0.71 (0.69 – 0.74) | 0.68 (0.64 – 0.71) | 0.80 (0.77 – 0.84) |
| Compressed 512x512 and 256x256 images without decompression | | | | |
| 512x512 | 0.71 (0.68 – 0.74) | 0.69 (0.66 – 0.72) | 0.70 (0.66 – 0.74) | 0.77 (0.73 – 0.81) |
| 256x256 | 0.68 (0.65 – 0.72 | 0.67 (0.65 – 0.70) | 0.61 (0.57 – 0.66) | 0.76 (0.72 – 0.80) |

Note – Numeric ranges in parenthesis represent 95% confidence intervals calculated bootstrapping based on random sampling 10 values at a time for 100 times

The customized ResNet50 architecture encompasses a pretrained ResNet50 model loaded with ImageNet weights, with these weights being frozen, excluding the Batch Normalization Layers. To tailor the model for our specific task, six fully trainable layers are appended, featuring 2048, 1024, 512, 256, 128, 32, and 1 layers of neurons. These layers are strategically added to enable the model to learn and extract pertinent features crucial for this classification objective.

- Input Layer : It receives three channels of Red, Green, and Blue (RGB) with a fixed resolution. These input images encompass three distinct types of breast images: the original source images sized at 1024 x 1024 pixels, decompressed images sized at 1024 x 1024 pixels obtained from the compression of 512 x 512-pixel images, and further compressed images to 256 x 256 pixels.

- Activation Function : Rectified Linear Unit (ReLU) activation function has been deliberately chosen. This choice is guided by prior research and practices within the field of Computer-Aided Diagnosis (CADx) in deep learning [21].

- Pooling Strategy : Pooling selects important features from the source feature map and creates smaller one. Average pooling computes the average pixel values of the pooling convolution [22].

- Output Layer : Guided by other medical binary classification tasks [23], sigmoid function has been chosen as the activation function for the last neuron.

To facilitate comprehensive comparisons, five distinct binary classification models have been trained. The first model is trained using the original EMBED images with a resolution of 1024x1024 pixels. The second model is trained using images decompressed from 512 x 512 pixels to 1024x1024 pixels. The third model utilizes images decompressed from 256x256 pixels to 1024x1024 pixels. The fourth and fifth models are trained by compressed images of 512x512 and 256x256 pixel images. These distinct models serve to assess and analyze the impact of image resolution and compression on the performance and efficacy of the breast cancer classifier, providing valuable insights into the model's robustness and suitability for varying image preprocessing approaches.

The optimal performing models from the three training datasets were selected based on the performance observed over 50 epochs of training. For gradient descent strategy, the first-order gradient-based optimization (ADAM) was chosen to determine the most suitable weights for neurons within the network. The ADAM optimizer is widely applied in numerous applications due to its efficiency and effectiveness in optimizing models [24]. In addition, binary cross-entropy was employed as a commonly utilized loss function for binary classification tasks, aiding in the assessment of model performance by quantifying the difference between predicted and actual outputs [25]. The determination of the optimal learning rate and other critical hyperparameters was accomplished through the Bayesian optimization scheme, a method known for efficiently optimizing model performance by intelligently exploring the hyperparameter space [26]. A total of three NVIDIA RTX 3090 GPUs with 31,488 CUDA cores, 984 Tensor Cores, and 72 GB VRAM, were utilized throughout the optimization process.

IV. RESULTS

The size of the source mammography dataset is 564MB, and the compressed dataset size with 512x512 resolution is 266MB. The compressed dataset size with 256x256 resolution is 81MB. But the ZIP compression of the source dataset is 577MB as shown in Table 3.

To assess the model's consistency in performance across test datasets, which represent unseen data, a bootstrapping analysis was carried out, involving random sampling of 10 images at a time from 493 images of 1024x1024 test datasets from the source, decompressed images of 1024x1024 from 512x512 and 256x256 resolutions. This sampling was repeated 100 times, providing valuable insights into models on test datasets.

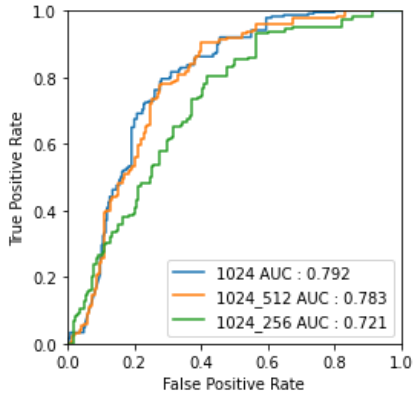

**Figure 2. Model Performance Comparisons between the Source Images and Decompressed Images on Test Datasets**

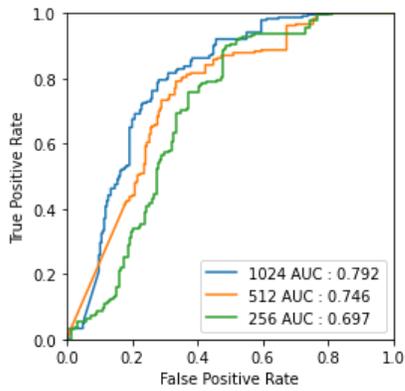

**Figure 3. Model Performance Comparisons between the Source Images and Compressed Images on Test Datasets**

The customized ResNet50, specifically tailored and fine-tuned on 1024x1024 mammograms that were decompressed from 512x512 resolution, demonstrates performance comparable to the model trained on the original 1024x1024 source images. Furthermore, the accuracy, precision, and recall of the customized model are 0.74 [0.72-0.77], 0.75 [0.70-0.79], and 0.72 [0.67-0.77], respectively, while the model trained on the source dataset showcases slightly higher values: accuracy of 0.75 [0.73-0.78], precision of 0.73 [0.69-0.77], and recall of 0.78 [0.75-0.82]. These results are summarized in Figure 2, and Table 4.

The customized ResNet50, trained using decompressed 1024x1024 images originating from 256x256 resolution, has not demonstrated identical performance with the one trained by the source images in comparison to the model trained on 1024x1024 images decompressed from 512x512 resolution. Both ResNet50 models trained on compressed images—specifically 512x512 and 256x256 images—exhibit noticeable performance degradation. Contrarily, utilizing decompressed images from 512x512 and 256x256 resolutions results in a 7% and 6% performance increase, respectively, in terms of Area-Under-Curve (AUC) when compared to the models trained on compressed images as shown in Figure 3 and Table 4.

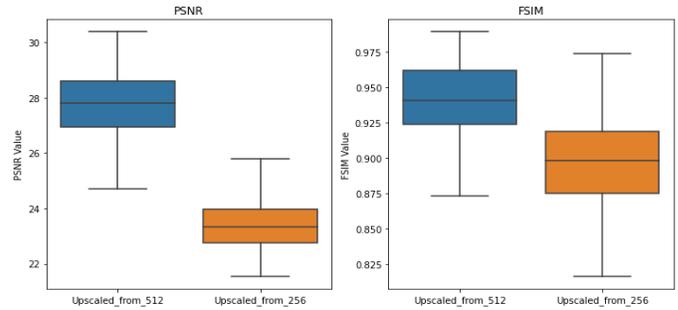

**Figure 4. PSNR and FSIM scores for Decompressed Images**

Image quality for the decompressed images has been assessed using conventional measures. When assessed by Feature Similarity Index Measure (FSIM) [27] metric, the decompressed 1024x1024 images from 512x512 images demonstrate a mean score of 0.93 with a variance of 0.0006. Similarly, applying the FSIM metric to the decompressed 1024x1024 images from 256x256 images yields a mean score of 0.89 with a variance of 0.001. These metrics signify a high degree of similarity, with the decompressed 1024x1024 images from 512x512 images being 93% identical to the source images. Additionally, evaluating the image quality using Peak Signal-to-Noise Ratio (PSNR) scores for the decompressed 1024x1024 images from 512x512 images reveals a mean score of 28.42dB with a variance of 8.32 dB. This PSNR score indicates an acceptable level of image quality, particularly in the context of lossy compression techniques [28]. These scores are depicted in Figure 4.

In the domain of medical image compression and decompression schemes, researchers have predominantly directed their focus towards enhancing wavelet transformations. However, this approach exhibits a significant drawback, notably the high consumption of computing resources. Our proposed scheme addresses this concern by mitigating the resource burden by employing downsampling operations, as outlined in [29]. Lossy compression algorithms, in general, contend with noise issues during image decompression. The DDPM-based compression and decompression framework presented herein are designed to effectively mitigate noise in decompressed images because training and fine-tuning of the DDPM-based decoder is expected to accurately computes the noise distribution in the decompressed images, subsequently eliminating it during the decompression process. Experimental results affirm the effectiveness of our compression and decompression framework, showcasing compression capabilities akin to lossless techniques while preserving essential image details and minimizing image quality degradation. Additionally, our compression performance surpasses that of ZIP lossless compression for image files.

## V. CONCLUSION

The compression and subsequent decompression of medical images using the proposed framework based on LDM and the Torchvision library demonstrated successful outcomes in our experiments. The customized ResNet50 models, trained using decompressed 1024x1024 images from 512x512, exhibit performance comparable to models trained on the original source images. It is noteworthy that this achievement was empowered by the particular "sd-x2-latent-upscaler" pretrained model as the basis for the framework. However, there exists a significant potential for further enhancement in the performance of the customized ResNet50 model, particularly the one more tailored towards decompressed 1024x1024 images from 256x256. Fine-tuning the 2x upscaler with EMBED mammograms may yield improvements in PSNR and FSIM scores, contributing to an overall enhancement in the final image quality and model performance. Such advancement would be a great contribution in the field by offering potential refinement in the compression and decompression framework.

The analysis of the proposed compression and decompression framework for medical images reveals their inherent merits and potential advantages upon finetuning LDM decoder. This compression framework is anticipated to fulfill the critical storage and bandwidth prerequisites for telemedicine services. Comprehensive evaluations encompassing qualitative and quantitative aspects of the proposed compression algorithm, such as image quality, deep learning performance, and compressed file sizes in comparison to ZIP compression, have been conducted. Experimental results affirm that our proposed framework functions effectively as a near lossless compression algorithm, preserving image quality while achieving a high compression rate.